%% file: main.tex
\pdfoutput=1
\documentclass[sigconf]{acmart}

\usepackage{booktabs} 
\usepackage{comment}
\usepackage{hyperref}
\usepackage[utf8x]{inputenc}
\usepackage[T1]{fontenc}
\usepackage{bm}
\usepackage{amsmath}
\usepackage{amsfonts}
\usepackage{amssymb}
\usepackage{comment}
\usepackage{graphicx}
\usepackage{multicol}
\usepackage{nomencl}
\usepackage{subcaption}

\usepackage{bbm}

\usepackage{amsthm}

\usepackage{hyperref}
\usepackage{graphicx}
\usepackage{subcaption}

\usepackage{booktabs}       
\usepackage{amsfonts}       
\usepackage{nicefrac}       
\usepackage{microtype}      
\usepackage{color}
\usepackage{amsmath}
\usepackage{amsthm}
\usepackage{natbib}
\usepackage{enumitem}
\usepackage{wrapfig}
\usepackage{lipsum}
\usepackage{multirow}
\usepackage{booktabs}
\usepackage{array}
\newcolumntype{P}[1]{>{\centering\arraybackslash}p{#1}}
\graphicspath{{figures/}}

\newtheorem*{lemma*}{Lemma}

\setcopyright{none}
\usepackage{tikz} 
\usetikzlibrary{positioning}
\usetikzlibrary{shapes,arrows} 
	\tikzstyle{decision} = [diamond, draw, fill=white!20,
	text width=4.5em, text badly centered, node distance=2cm, inner sep=0pt]
	\tikzstyle{block} = [rectangle, draw, fill=white!20,
	text width=15em, text centered, rounded corners, node distance=3cm, minimum height=2.5em]
	\tikzstyle{smallblock} = [rectangle, draw, fill=white!20,
	text width=5em, text centered, rounded corners, node distance=3cm, minimum height=2.5em]
	\tikzstyle{line} = [draw, -latex']
	\tikzstyle{cloud} = [draw, ellipse,fill=white!20,  node distance=2cm,
	minimum height=2em]

\acmConference[e-Energy'19]{the Tenth International Conference on Future Energy Systems}{June 2019}{Phoenix, AZ, United States}
\acmYear{2019}
\copyrightyear{2019}

\begin{document}
\title{Exploiting Vulnerabilities of Load Forecasting Through Adversarial Attacks}

\author{Yize Chen}
\affiliation{%
  \institution{Electrical and Computer Engineering, \\University of Washington}
  \city{Seattle}
  \state{WA}
  \postcode{98195}
}
\email{yizechen@uw.edu}

\author{Yushi Tan}
\affiliation{%
  \institution{Electrical and Computer Engineering, \\
  	University of Washington}
  \city{Seattle}
  \state{WA}
  \postcode{98195}
}
\email{ystan@uw.edu}

\author{Baosen Zhang}
\affiliation{%
  \institution{Electrical and Computer Engineering, \\University of Washington}
  \city{Seattle}
  \state{WA}
  \postcode{98195}
}
\email{zhangbao@uw.edu}



\begin{abstract}
\input{abstract}
\end{abstract}


\maketitle

\section{Introduction}
\label{sec:intro}
\input{intro}

\section{Related Work}
\label{sec:related}
\input{relatedwork}

\section{Formulation: Forecasters and Attackers}
\label{sec:formulation}
\input{formulation}

\section{Blind Attack on Load Forecasting}
\label{sec:method}
\input{method}

\section{Attacks on System Operations}
\label{sec:market}
\input{system}

\section{Case Studies}
\label{sec:simulation}
\input{simulation}

\section{Discussion and Conclusion}
\label{sec:conclusion}
\input{conclusion}

\bibliographystyle{ACM-Reference-Format}
\bibliography{ref}
\appendix
\input{appendix}

\end{document}

%% file: abstract.tex
Load forecasting plays a critical role in the operation and planning of power systems. By using input features such as historical loads and weather forecasts, system operators and utilities build forecast models to guide decision making in commitment and dispatch. As the forecasting techniques becomes more sophisticated, however, they also become more vulnerable to cybersecurity threats. In this paper, we study the vulnerability of a class of load forecasting algorithms and analyze the potential impact on the power system operations, such as load shedding and increased dispatch costs. Specifically, we propose data injection attack algorithms that require minimal assumptions on the ability of the adversary. The attacker does not need to have knowledge about the load forecasting model or the underlying power system. Surprisingly, our results indicate that standard load forecasting algorithms are quite vulnerable to the designed black-box attacks. By only injecting malicious data in temperature from online weather forecast APIs, an attacker could manipulate load forecasts in arbitrary directions and cause significant and targeted damages to system operations. 

%% file: intro.tex
Load forecasting is a fundamental step in power system planning and operations. It is used to inform system operators the future load profiles, and serves as the basis of decision-making problems such as unit commitment, reserve management, economic dispatch and maintenance scheduling~\cite{gross1987short}. Consequently, the accuracy of forecasted loads directly impact the cost and reliability of system operations~\cite{hobbs1999analysis}. 
With a growing penetration of new technologies into the demand side, the utilities and system operators need to place more importance on both accurate and robust forecasts.

For years, the holy grail in short-term load forecasting has been to \emph{improve the forecast accuracy}, which has been vigorously pursued by the research community. The variations in load are driven by many different factors, including temperature, weather, temporal and seasonal effects (e.g., weekday vs. weekend) and other socioeconomic factors. All of these factors influence the load in nonlinear and complex ways. 
 Over the past decades, a myriad of load forecasting algorithms have been proposed and adopted in practice. See, for example, \cite{gross1987short, de200625, charytoniuk1998nonparametric} and the references within.

For simplicity, in this paper, we restrict the inputs of the algorithms to be the historical load data, time indicators and temperature information. These algorithms can be thought of as finding a mapping between the (high dimensional) input features to the forecasted time series of load values. Statistical and machine learning techniques, such as support vector regression~\cite{ceperic2013strategy}, ARIMA~\cite{contreras2003arima} and neural networks~\cite{hippert2001neural, chen2018short} have been applied to short term load forecasting and are well adopted in practice. The recent advances in deep learning opened the door to using more input features and deeper model architectures to further improve load forecasting accuracy and provided some of the best performances to date~\cite{kong2017short, quilumba2015using,wang2016electric}.

\begin{figure}[h]
	\centering
	\includegraphics[width=1.0 \columnwidth]{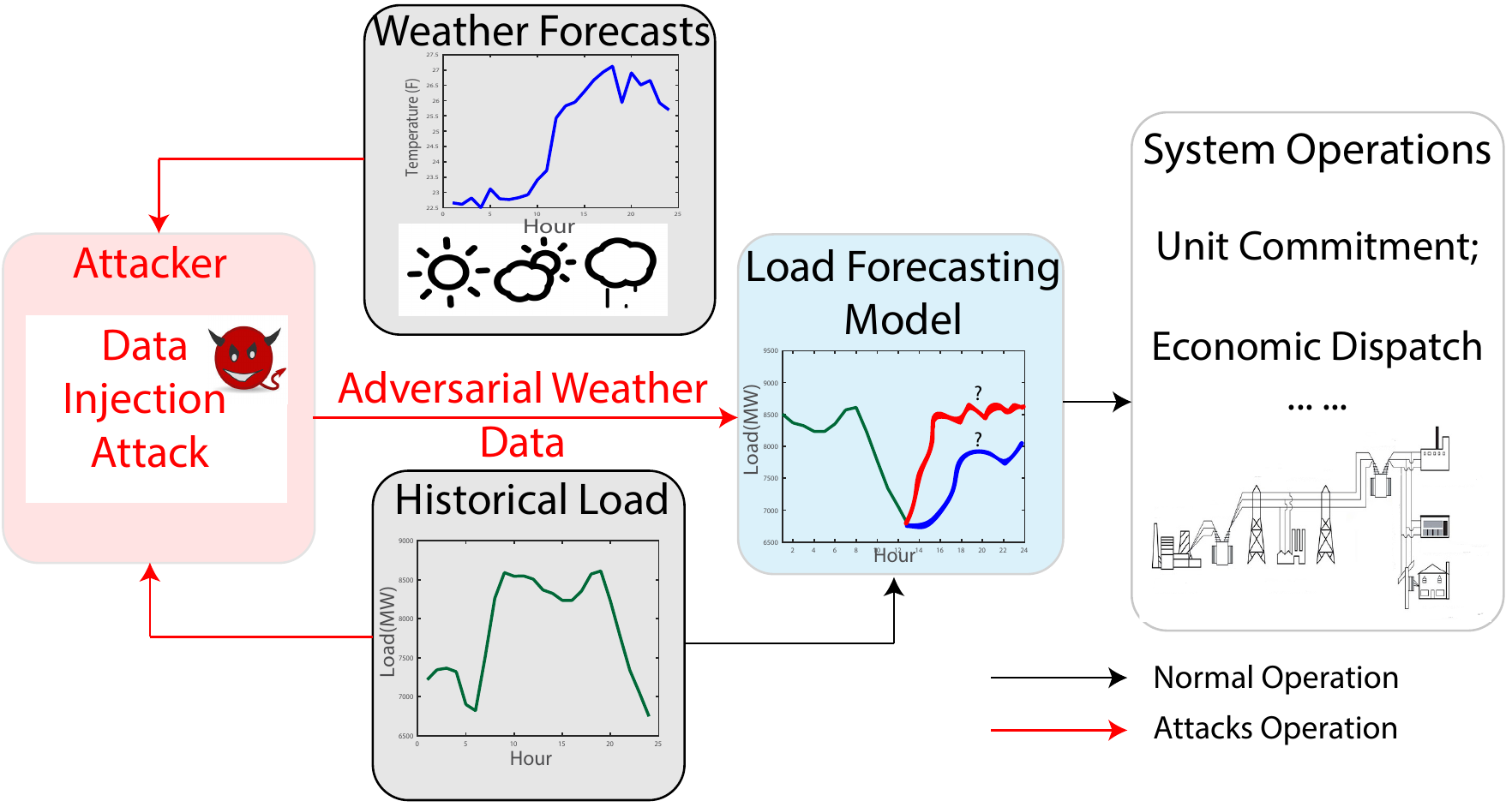}
	\caption{The schematic of our proposed attacks on load forecasting algorithms along with the threats over power system operations. Without knowledge about the forecast model's parameters, the attacker injects designed small, undetectable data perturbations into weather forecasts to induce abnormal system operations.} 
	\label{fig:schematic}
\end{figure}

As the forecasting methods become more complex and accurate, they are also more susceptible to cybersecurity threats. In this paper, we look into the data vulnerabilities of such methods, where an attacker adversarially injects false data into the input features of forecasting algorithms. Specifically, we investigate false data injection attacks of the temperature data. It is an important input to load forecasting algorithms and is mostly obtained from external services/APIs, therefore providing an easier avenue for data perturbations and attack injections. The potential damage of these types of attacks can be significant, leading to increases in system operation costs and maybe even more catastrophic events such as load shedding. 
In Figure~\ref{fig:schematic}, we show the schematic of threats and proposed attacks to systems.

In this paper, we take the perspective of an attacker and develop attack strategies on load forecasting algorithms, and conduct damage analysis of the proposed attacks. We take a restrictive setting of both the attacker's ``knowledge'' and ``capabilities'', where the attacker does not know any parameter of the targeted load forecasting algorithms, and could only inject perturbations into input temperatures under constraints to avoid detection. 

Under this setup, we develop two simple data-driven attack strategies for finding the injected perturbations onto the original temperature data. Surprisingly, we find the proposed attacks significantly degrade the performance of a class of (accurate) load forecasting algorithms. With only few degrees of perturbations injected into input temperatures, the load forecasting algorithm's output deviates drastically from original values. We also assess the damages brought by such model vulnerabilities in power system operations. Simulations based on real-world load datasets show that by changing only few degrees of temperature, adversarial forecasts not only increase the operation cost of power systems, but can also lead to load shedding and infeasible generator schedules. 

This study illustrates the need to look at other properties of load forecasting techniques in addition to \emph{forecast accuracy}. We demonstrate that accuracy may not mean robustness, and since a wrong forecast of load potentially leads to costly operation decisions or system damages, we call for a more comprehensive analysis when developing and applying load forecasting techniques. Specifically, we make the following contributions in this work:
\begin{itemize}
	\item To the best of our knowledge, this is the first to evaluate the security issues of load forecasting procedures in power system operations. Data vulnerabilities of current forecasting methods are  discussed and formulated.
	\item Two data-driven, black-box attack algorithms, namely \texttt{learn and attack} and \texttt{gradient estimation}, are proposed to generate hard-to-detect, adversarial input data for load forecasting algorithms.
	\item Case studies on power system operations demonstrate potential damages via proposed attacks. We show that the strategically designed adversarial injections could lead to either increased system operating costs or load shedding.
\end{itemize}

We make our code open source on load forecasting model development, attack implementations and market operation evaluation, and make it as a package for evaluating load forecasting robustness and security\footnote{\url{https://github.com/chennnnnyize/load_forecasts_attack}}.

The rest of the paper is organized as follows. A literature review is presented in Section~\ref{sec:related}; we then briefly summarize a general load forecasting model, and formulate the objective and constraints of attackers in Section~\ref{sec:formulation}; in Section~\ref{sec:method}, we detail the algorithms for implementing the attack; to illustrate the attack's threats to the power system operations, we describe the market setup and a toy example in Section \ref{sec:market}; through simulations based on real-world load data in Section~\ref{sec:simulation}, we demonstrate the threats posed by the proposed attacks; furthur discussion on model/data security and conclusion are drawn in Section~\ref{sec:conclusion}.

%% file: relatedwork.tex
In this section, we give brief literature review on both the load forecasting methods and cyber-security of power systems. Our work is different from most related work in two aspects: most of the studies in forecasts do not consider security and robustness, while  most of the studies in power system security evaluate attacks with almost no knowledge about the targeted system or constrained capabilities.

Our work is related to the large body of work on forecasting in power networks, such as renewables forecasting~\cite{pinson2007trading} and load forecasting~\cite{gross1987short, park1991electric}. Since the costs of making erroneous forecasts are so high, even reducing forecast error in a few percent points are important~\cite{de200625}. Various methods have been applied and evaluated in load forecasting problems, including using nonparametric regression~\cite{charytoniuk1998nonparametric}, support vector regression~\cite{ceperic2013strategy}, ARIMA~\cite{contreras2003arima} and neural networks~\cite{hippert2001neural, chen2010short}. Among these forecast models, neural network has become increasingly more popular, as it provides highly accurate results due to the ability of  representing the complex relations between high-dimensional features and outputs. 

The recent progress in deep learning and data science also promotes the use of deep neural networks and more complicated feature representations in forecast models~\cite{chen2018short, kong2017short, chen2017modeling}. Many works focus on feature selection and feature engineering by considering the uncertainties coming from both electrical loads~\cite{wang2018data} and exogenous variables such as weather~\cite{hong2010cost, wang2016electric}, customer behaviors~\cite{quilumba2015using} etc. However, most research doesn't look into the robustness issues, and model performances under  adversarial environements are rarely discussed~\cite{luo2018benchmarking, chen2018machine}. 

Our work is also under the scope of cyber-physcial system security, especially the cyber-security of power systems~\cite{mcdaniel2009security}. Many studies focused on compromising the communication, sensing or monitoring process in modern smart grids~\cite{sridhar2012cyber, mo2009secure}. For instance, \emph{denial of service attacks} and \emph{deception attacks} are aimed at compromising either communication channel or communication packets~\cite{amin2009safe}; false data injections on state estimation have been widely discussed~\cite{kosut2010malicious, liu2011false}, where the attackers introduce estimation errors on state variables, e.g.,  phase angles and voltage magnitudes. Such attacks strategically manipulate meter measuresments to bypass conventional bad data detection.  In~\cite{xie2010false,tan2018online}, the authors analyzed how maliciously changed system states could affect the market operations during dispatch process.  Most of the previous attacks assume full knowledge of system configuration. 
It is also assumed that attackers possess strong capabilities to implement attacks, e.g., to compromise communication channel or to modify meter data arbitrarily.

In this paper, we focus on the previously overlooked vulnerabilities in the load forecasting process. For instance, forecasting model inputs can be exposed to adversarial modification and the model performance may be impacted by such malicious changes. Recently, there has been a hot debate on the security of machine learning models~\cite{szegedy2013intriguing} following the deep learning's state-of-the-art achievements on a bunch of benchmark tasks. 
In computer vision, researchers found a small, adversarially designed noises injected to clean image would deceive a well-trained image classifier~\cite{papernot2016limitations, bhagoji2018practical}. We are interested in whether such attacks could also impact the performance of load forecasting models and if so to what extent. The proposed class of data injection attacks do not assume the forecasting model itself is known to the attackers.  In addition, successful distortion on load forecasting also impacts the reliable operations of power systems, so it is important to investigate the data vulnerabilities in existing load forecasting methods.

%% file: formulation.tex
In this section, we formally describe the forecasting and attacking models. To set up realistic vulnerability analyses, we also describe the set of restrictions on the knowledge and capability of the attacker.

\subsection{Load Forecasting Formulation}
The schematic of general load forecasting model is depicted in Figure~\ref{fig:schematic}. We consider the setup for a family of load forecasting algorithms with different architectures. The input features of these algorithms include historical records of load, weather forecasts including temperature, weather indicators (e.g., sunny, rainy or cloudy) and seasonal indicator variables such as weekdays/weekends and hour of the day. 
Mathematically, the system operator would be able to collect a training dataset $\mathcal{D}_{tr}=\{(\textbf{X}_{t-H},..., \textbf{X}_t);L_{t+k}\}_{t=1}^T$ based on available historical data. Here $L_{t+k}\in [0,1]$ are scalars representing scaled load values (or \emph{response variables})~\cite{gross1987short}.  $k$ is the model's forecast horizon, typically ranging from one hour to one day in short-term load forecasts. $\textbf{X}_{t-i}\in [0,1]^d,\; i=0,...,H$ are scaled, $d$-dimensional input feature verctors (or \emph{numerical predictor variables}). Let's denote $\textbf{X}_t:=\{L_t, \textbf{X}_t^{temp}, \textbf{X}_t^{index}\}$, where $L_t$ are the load history records;  $\textbf{X}_t^{temp}$ are the temperature value vectors, which could be acquired from either system historical records or weather forecast API; $\textbf{X}_t^{index}$ are a collection of indicators, indicating the weather characteristics, seasonal factors and time factors. $H$ determines how much history of training data the operators want to take into consideration for forecasting. Longer history would provide more information to the forecast model, yet brings more difficulty in model training and fitting. 

In the task of load forecasting, one is interested to find a function parameterized by $\theta$: $f_\theta(\textbf{X}_{t-H}, ..., \textbf{X}_t)=\hat{L}_{t+k}$, which learns the mapping from $(\textbf{X}_{t-H},...,\textbf{X}_t)$ to future loads $\hat{L}_{t+k}$.  The mean absolute error~(MAE) is widely used to measure the performance of forecasting algorithm, which is defined by the  average $L_1$ norm of difference on forecasted loads. Estimation of $\theta$ is given by minimizing the $L_1$-norm of the difference between model predictions and ground truth values:
\begin{subequations}
	\begin{align}
		\min_\theta \quad &\frac{1}{T}\sum_{t=1}^{T}||f_{\theta}(\textbf{X}_{t-H},...,\textbf{X}_t)-L_{t+k}||_1 \\
		s.t.  \quad  & \theta \in \Theta
	\end{align}
	\label{equ:forecasts}
	\vspace{-8pt}
\end{subequations} 

During training, ground truth of historical records on $\textbf{X}_t$ and $L_{t+k}$  are used; during testing and real-world system implementations, we are using $\textbf{X}_t$ which are coming from weather forecasts to forecast future loads. Once the model is learned, it can be applied in a rolling-horizon fashion to make use of forecasted $\hat{L}$ along with $\textbf{X}_t^{temp}$ and $\textbf{X}_t^{index}$ to forecast for furthur into the future.

\subsection{Specific Forecasting Models}
We describe the model setup for several representative load forecasting algorithms which have achieved good performances and have been widely adopted~\cite{hippert2001neural}. In Appendix \ref{sec:app} we detaild the model parameter settings and training approaches. We note that the vulnerabilitiy analysis conducted by this paper is not constrained to the following forecasting algorithms. As long as the model output is sensitive with respect to input features, our proposed attack methods would be able to alter the load patterns maliciously.

\subsubsection{Feed-Forward Neural Networks}
A multi-layered, feed-forward neural networks~(NN) has been widely used to represent the nonlinearities between input features and output forecasts~\cite{hippert2001neural}. For the input layer of neural networks, each neuron represents one feature of training input, and all features of past $H$ steps $(\textbf{X}_{t-H},...,\textbf{X}_t)$ are stacked as the inputs. For each intermediate layer, NN could have a tunable number of hidden units, which represent the input feature combinations. Recent advances in deep learning also allow for deeper and more comlicated network design~\cite{chen2018short}.

\subsubsection{Recurrent Neural Networks}
A recurrent neural networks~(RNN) is a  class of neural networks that are specially designed for sequential modeling~\cite{vermaak1998recurrent}. Instead of stacking all time steps' features together as in the feed-forward neural networks, RNN feeds each step's input $\textbf{X}_t$ sequentially, and outputs a hidden unit to represent the feature combination of current input and historical features. The last neuron outputs the forecasted load values.

\subsubsection{Long Short-Term Memory}
Long Short-Term Memory network~(LSTM) is designed to deal with the vanishing gradient problem existing in the RNN with long-time dependencies~\cite{kong2017short}. The major improvements over RNN are the design of ``forget" gates to model the temporal dependencies and capture long time dependencies in load patterns more accurately.

\subsection{Objective of Attacker}
The attacker's goal is to distort the forecasted load as much as possible in a certain direction, e.g., to either increase or decrease forecasted values. In order to distort the output forecast values, the attacker actually has two choices of inserting attacks: \emph{to attack $\textbf{X}_t$} or \emph{to attack $\theta$}. While the trained model itself is often safely kept by the operators, it has to use external data such as weather forecasts $\textbf{X}_t^{temp}$ as input features. Then the attacker's goal is to inject perturbations into the weather forecasts coming from external services to generate adversarial input data $\tilde{\textbf{X}}_t^{temp}$ for $f_\theta(\cdot)$, so that predictions are modified. We use $\gamma=\{-1,1\}$ to denote the chosen attack direction by attackers. If $\gamma=1$, the attacker tries to find $\tilde{\textbf{X}}$ to decrease the load forecast values; when $\gamma=-1$, the attacker tries to find $\tilde{\textbf{X}}$ to  increase load forecasts values. Since load values are always positive, the attacker's goal is to  find $\tilde{\textbf{X}}$ that minimizes the value of $\gamma f_{\theta}(\tilde{\textbf{X}}_{t-H},...,\tilde{\textbf{X}}_t)$.

\subsection{Attacker's Knowledge}
We consider two attack scenarios, \emph{white box} and \emph{black-box} attacks. 
In the \emph{white-box} settings, the attacker is assumed to know exactly the model parameters $\theta$. This is a strong assumption in the sense that load forecast model $f_\theta(\cdot)$ is fully exposed to the attacker. On the contrary, in the \emph{black-box} setting, the attacker only knows which family of load forecasting model has been applied~(e.g., NN or RNN), but is blind to the forecasting algorithms and has no knowledge of any parameters of $f_\theta$. We consider two possible avenues of attacks. In the first case, the attacker possesses a \emph{substitute training dataset} $\mathcal{D}_{tr}'$ which may or may not be the same as $\mathcal{D}_{tr}$. Such dataset also represents the ground truth of historical load and features. In the second case, the attacker cannot acquire such dataset due to lack of access to the historical load records. We assume the attacker has \emph{query} access to the load forecasting model\footnote{Such query access assumption is possible in many \emph{forecast-as-a-Service} businesses, e.g., SAS energy forecasting and Itron forecasting.}.  That is, the attacker could query the implemented load forecasting model by using different values of input features for a limited number of times, and then try to get insights on how $f_\theta$ works.

\begin{figure}[h]
	\centering
	\includegraphics[width=1.0 \columnwidth]{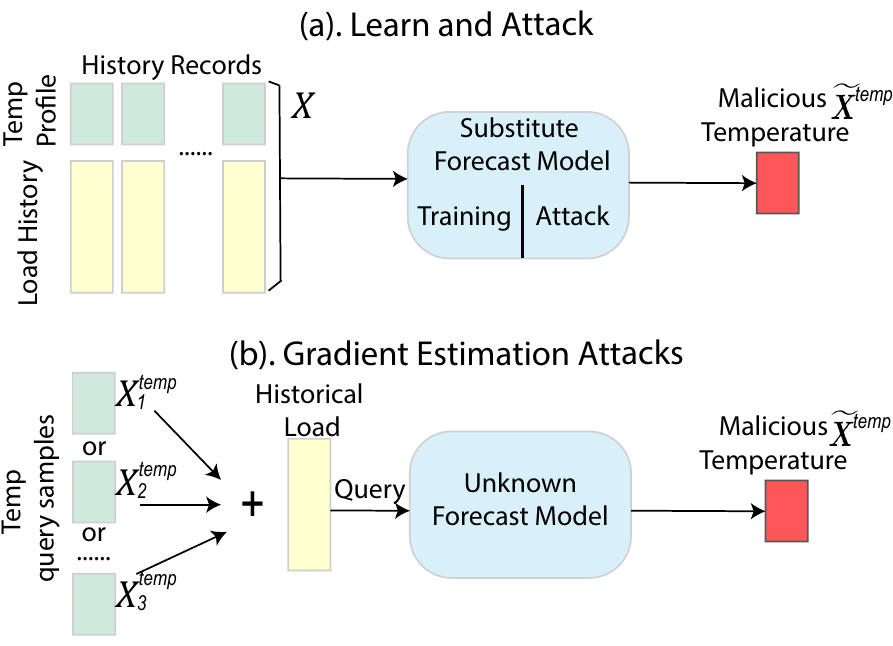}
	\caption{Schematics for two proposed attacks on load forecasting models by changing input temperature vectors: (a). the \texttt{learn-and-attack} algorithm and (b). the \texttt{gradient estimation} algorithm.}
	\label{fig:loadforecasting}
\end{figure}

\subsection{Attacker's Capability}
As an attacker, it is important to avoid being detected by the bad data detection algorithms used by system operators. The attacker's capability could be upper bounded by the allowed number of perturbed entries in the input data; it could be bounded by the average deviations on all features; or it could be also bounded by the largest deviation from the original value. Mathematically, the attacker wants to keep $||\tilde{\textbf{X}}_t^{temp}-\textbf{X}_t^{temp}||_p$ 
bounded, where $p$ can take different values such as $0, 1, \infty$ to express certain norm constraints corresponding to different detection algorithms.

In summary, we formulate the model of attackers as the following optimization problem:

\begin{subequations}
	\begin{align}
	\label{equ:obj}
	\min_{\tilde{\textbf{X}}_{t-H}^{temp},...,\tilde{\textbf{X}}_t^{temp}}  \quad & \gamma f_{\theta}(\tilde{\textbf{X}}_{t-H},...,\tilde{\textbf{X}}_t)\\
	\label{equ:constraint}
	s.t.  \quad  & ||\textbf{X}_{t-i}^{temp}-\tilde{\textbf{X}}_{t-i}^{temp}||_p \leq \epsilon,\; i=0,...,H
	\end{align}
	\label{equ:attack}
\end{subequations} 

\begin{figure*}[h]
	\centering
	\includegraphics[width=2.1 \columnwidth]{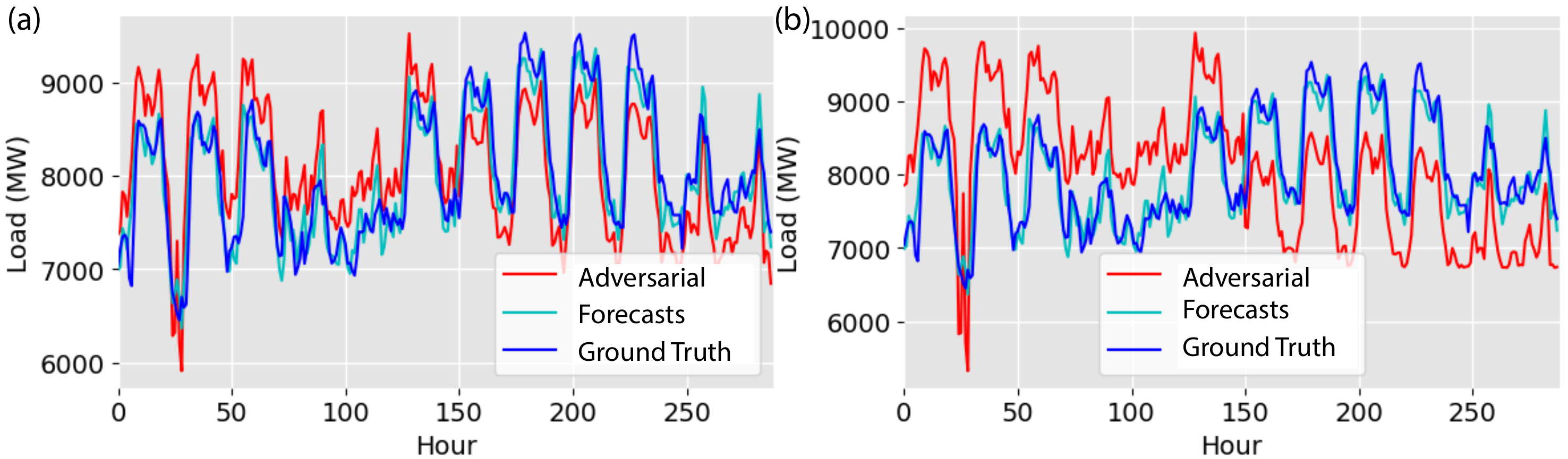}
	\caption{We show $300$ hours forecasts based on original and adversarial temperature data for the aggregated load of Switzerland. The load forecasting algorithn is an recurrent neural networks with inputs composed of past load, regional temperature forecast values and weather indicators.  The attack perturbations are generated by using the \texttt{learn and attack} method, and it implements load maximization strategy in the first $150$ hours and load minimization strategy in the latter $150$ hours. (a). Load forecasting results with  temperature attack constraint of (maximum perturbations) $1F$; (b). load forecasting results with temperature attack constraint of $5F$.}
	\label{fig:forecast_example}
\end{figure*}

Note that there is a parallel between the forecast problem \eqref{equ:forecasts} and attack problem \eqref{equ:attack}, where the objective's optimization directions and optimization variables are exactly in the opposite directions: forecasting model works on model parameters to minimize forecast errors, while attacker works on model inputs to maximize the errors to targeted directions. However, due to lack of model knowledge in the black box setting, it is a challenging task for attackers to find efficient attack input $\tilde{\textbf{X}}^{temp}$ via \eqref{equ:attack}. In the next section, we will show two attack methods generally working with attacker's knowledge coming from \emph{substitute training dataset} and \emph{query access} respectively.

%% file: method.tex
In this section, we first describe attacks under the \emph{white-box} setting, where an attacker possesses full knowledge of load forecasting model parameters. This serves as a benchmark for evaluation of the success of attackers. We then focus on two more realistic settings where the attacker does not know the model parameters. We describe how data injection attacks can be implemented when either the historical data is known or the attacker has limited query access to the load forecasting model.

\subsection{White-Box Attack}
Under the \emph{white-box} setting, since the model parameters are known to the attacker, it is possible to find the attack input via solving \eqref{equ:attack}. For the convenience of notations, we omit the superscript on $\textbf{X}$ in some of the following paragraphs, and introduce the generalizable attack methods not only suitable for attacking temperature forecasts, but also suitable for injecting false data into other features.

Since most state-of-the-art load forecasting algorithms use complex models such as neural networks, the attacker's problem \eqref{equ:attack} is nonconvex and furthurmore, there is no closed-form solution for $\tilde{\textbf{X}}_{t-H},...,\tilde{\textbf{X}}_t$. Nevertheless, an attacker can still find some attack vectors iteratively by taking gradients with respect to each time step's temperature values. Even though this may not find the optimal solution to \eqref{equ:attack}, because of the highly nonconvex nature of the forecasting model, a slight (suboptimal) perturbation of the input features would drastically change the forecast output.

Based on \eqref{equ:attack}, we define a loss function $\mathcal{L}$ with respect to each time step's feature $\tilde{\textbf{X}}_{t-i}, \; i=0,...,H$. Then the attacker iteratively takes gradients of $\mathcal{L}$ to find the adversarial input $\tilde{\textbf{X}}_{t-i}$. The constraints in \eqref{equ:constraint} is included in the loss function using a log-barrier:
\begin{equation}
\label{equ:loss}
\mathcal{L}(\tilde{\textbf{X}}_{t-i})=\gamma  f_{\theta}(\tilde{\textbf{X}}_{t-H},...,\tilde{\textbf{X}}_t)- \beta \log( \epsilon- ||\textbf{X}_{t-i}^{temp}-\tilde{\textbf{X}}_{t-i}^{temp}||_p)
\end{equation}
where $\beta$ is the weight of the barrier term.
Since there are a large number of parameters and input features in many load forecasting algorithms, it can be computationally expensive to compute the exact gradient values for each input feature. We follow a simpler method in~\cite{szegedy2013intriguing} to only update the feature values based on the sign of the gradient at each iteration $j$:
\begin{equation}
\label{equ:grad_attack}
\tilde{\textbf{X}}_{t-i}^{(j+1)}=\tilde{\textbf{X}}_{t-i}^{(j)}-\alpha \cdot \text{sign}(\triangledown_{\tilde{\textbf{X}}_{t-i}^{(j)}} (\mathcal{L}(\tilde{\textbf{X}}_{t-i}^{(j)})))
\end{equation}
where $\alpha$ controls the step size for updating adversarial temperature values. The resulting adversarial temperature vector is obtained by applying \eqref{equ:grad_attack} a number of times.

\subsection{Learn and Attack}
In the \texttt{learn and attack} setting,  we assume the attacker does not have access to the model parameters, and there is no query access to the model. The only knowledge the attacker has is a historical dataset $\tilde{\mathcal{D}}_{tr}$, which includes same features as data set  $\mathcal{D}_{tr}$ used to train the load forecasting model\footnote{In Learn and Attack setting, we make assumption that the attacker know the family of targeted load forecasting model,  e.g., a feedforward neural networks or a Recurrent Neural Networks.}. The proposed attack algorithm consists of a \emph{training phase} and an \emph{attack phase} as shown in Fig.~\ref{fig:loadforecasting}(a). In the training phase, the attacker trains substitute model $f_{\tilde{\theta}}$ based on $\tilde{\mathcal{D}}_{tr}$ to minimize the training loss.  In the attack phase, the attacker pretends that the substitute model is the true load forecast model and performs white-box attacks on it to find the attack vectors. This strategy is based on the assumption that the substitute model behaves similarly to the true model not only for the training data $\textbf{X}$, but also for the attack vector $\tilde{\textbf{X}}$. Then by injecting $\tilde{\textbf{X}}$ into the true load forecasting model, the forecast values go to attacker's desired directions.

It is useful to evaluate the \emph{transferability} of proposed attacks across different set of models $f_\theta$ and $f_{\tilde{\theta}}$. The phenomenon of transferability in adversarial attacks for machine learning models have been discussed in  \cite{papernot2016transferability, hosseini2017blocking}, where adversarial instance generated using $f_{\tilde{\theta}}$ can be also treated as an adversarial instance by $f_{\theta}$ with high probability. The theoretical understanding of why attacks transfer remains an open question and is out of scope for this paper. In Fig.~\ref{fig:forecast_example} we show such \emph{transferability} also exists in the load forecasting model. The temperature inputs are generated by implementing the iterative gradient update \eqref{equ:grad_attack} based on a substitute model under $L_\infty$-norm of attack perturbations, yet such adversarial temperature values also mislead the (unknown) true load forecasting model to be wildly inaccurate.

\subsection{Gradient Estimation Attack}
To implement \texttt{learn and attack} on load forecasting algorithms, the attacker needs get a version of the training data to learn a substitute load forecasting model. In the case there is no available historical data records, if the attacker is able to query the load forecasting algorithm for a limited number of times, it is still possible to construct adversarial temperature inputs by using queries to estimate the gradients. In Figure \ref{fig:loadforecasting} (b) we show the schematic on generating adversarial temperature instances via querying.

For $k$-th dimension of the input feature at time stamp $t-i$, $\tilde{\textbf{X}}_{k,t-i}^{j+1}$, the attacker needs to query the load forecasting system on each feature dimension to calculate the two-sided estimation of the gradient of $f_\theta$:
\begin{equation}
\triangledown_{\tilde{\textbf{X}}_{k,t-i}}f_\theta(\tilde{\textbf{X}})   \approx \frac{f_\theta(\tilde{\textbf{X}}+\delta\mathbf{e}_k)-f_\theta(\tilde{\textbf{X}}-\delta\mathbf{e}_k)}{2\delta}
\end{equation}
where $\mathbf{e}_k$ is a $d$-dimensional vector with all zero except $1$ at $k$-th component, and $\delta$ takes a small value for gradient estimation. Once the gradient is estimated for each dimension of temperature features, we can follow the same method of \eqref{equ:grad_attack} to iteratively build the adversarial features using the estimated gradient vectors:
\begin{equation}
\label{equ:grad_attack2}
\tilde{\textbf{X}}_{t-i}^{(j+1)}=\tilde{\textbf{X}}_{t-i}^{(j)}-\alpha \gamma \cdot \text{sign}(\triangledown_{\tilde{\textbf{X}}_{t-i}}f_\theta(\tilde{\textbf{X}}^{(j)}) )
\end{equation}

To satisfy the norm constraints on the allowed perturbation of $\tilde{\textbf{X}}$, the attacker projects the adversarial data back into the pre-defined norms after finishing the iterative attack constructions. In~\cite{bhagoji2018practical} techniques on reducing number of queries are also discussed for attacking an image classifier, which could also help improve the query efficiency of load forecasting attacks.

%% file: system.tex
In this section, we first illustrate a power system operation case consisting of a day-ahead planning stage and a real-time operational stage, which is simple yet close to real-world market operations. We then describe two simple temporal attack strategies that pose threats to such system operations via injecting perturbations into load forecasting inputs.

\subsection{Power System Operations Model}

\begin{enumerate}
	\item 
	A commitment schedule based on the day-ahead load forecasts is created by a unit commitment (UC) model based on the day-ahead load forecast:
\begin{subequations}
	\label{equ:UC}
	\begin{align}
		\underset{\mathbf{u}, \mathbf{p} }{\min} \;\, & C(\mathbf{p}) + S(\mathbf{u}) \\
		\text{s.t.} & \sum_{g \in G} p_{g}^{t} = \hat{L}_t, \;\, \forall t \in T \label{eqn:UC_balance} \\
		& u_i^t p^{\text{min}}_g \leq p_g^t \leq u_g^t p^{\text{max}}_i, \;\, \forall g \in G, \;\, \forall t \in T \label{eqn:UC_GLim} \\
		& u_g^t - u_g^{t - 1} = z_g^t - y_g^t, \;\, \forall g \in G, \;\, t \in T \label{eqn:UC_GLogic} \\
		& \sum_{\tau = t - t_g^{\text{up}} + 1}^t z_g^{\tau} \leq x_g^t, \;\, \forall g \in G, \forall t \in T \label{eqn:UC_minup} \\
		& \sum_{\tau = t - t_g^{\text{dn}} + 1}^t z_g^{\tau} \leq 1 - x_g^t, \;\, \forall g \in G, \forall t \in T \label{eqn:UC_mindn} \\
		& -R_g^{\text{dn}} \leq p_g^{t + 1} - p_g^t \leq R_g^{\text{up}}, \;\, \forall g \in G \label{eqn:UC_Ramp} \\
		& u_g^t, z_g^t, y_g^t \in \{0, 1\}, \;\, \forall g \in G, \;\, t \in T
	\end{align}
\end{subequations}
	where $u_{g}^{t}$ is the binary decision variable of the commitment status of generator $g$ at time $t$, with $1$ indicating $g$ is online; $p_g^t$ is the real power output of generator $g$ at time $t$; all the $u_g^{t}$'s and $p_g^t$'s are collected together into vectors $\mathbf{u}$ and $\mathbf{p}$; $C(\mathbf{p})$ and $S(\textbf{u})$ represent the dispatch costs and startup and shutdown costs, respectively, of all the generators in all periods; the constraints are system-wide power balance constraint~(\ref{eqn:UC_balance}), generation limits constraints~(\ref{eqn:UC_GLim}), generator logical constraint~(\ref{eqn:UC_GLogic}), minimum up time constraint~(\ref{eqn:UC_minup}), minimum down time constraint~(\ref{eqn:UC_mindn}) and ramping constraints~(\ref{eqn:UC_Ramp}). Once solved, the operator gets the schedule for the set of online generators $G_t$ at each time $t$.
	
	\item For each time stage $t$ of each day, the dispatch of the scheduled units and the actual dispatch cost are calculated according to a basic Economic Dispatch (ED) model~\cite{kirschen2018fundamentals} based on the actual load and generation schedule $G_t$:
\begin{subequations}
	\begin{align}
	    \underset{ \mathbf{p}_t }{\min} \,\; & C(\mathbf{p}_t) \\
		\text{s.t.} & \sum_{g \in G_t} p_{g}^{t} = L_t, \;\, \label{eqn:ED_balance} \\
		& p^{\text{min}}_g \leq p_g^t \leq p^{\text{max}}_g, \;\, \forall g \in G_t \label{eqn:ED_GLim}
	\end{align}
\end{subequations}
	where it aims to find the real power dispatch at time $t$, $\mathbf{p}_t$, that minimizes the dispatch costs at time $t$, $C(\mathbf{p}_t)$, considering system-wide power balance constraint~(\ref{eqn:ED_balance}) and generation limits constraints~(\ref{eqn:ED_GLim}).
	The daily operation cost is obtained by summing the 24-hour dispatch costs and the startup and shutdown costs. When the ED based on the day-ahead commitment does not have a feasible solution, a load is shed to maintain the balance between supply and demand. 
\end{enumerate}

\subsection{Attack Strategies}
Under normal operating conditions, the load forecasting algorithms provide accurate forecasts on day-ahead load for system operators to solve~\eqref{equ:UC}. During an attack, adversarial temperature forecasts are injected into the day-ahead planning stage to cause deviation from the normal operations, e.g., increased system costs, load shedding, no feasible generation dispatch or violation of ramping constraints. We assume the attacker does not know the parameters of underlying system such as each generator's capacity and ramp constraints.

We propose two intuitive attack strategies that move the load forecasts as far away as possible to stress the system. Simple as it is, the toy example in Section~\ref{sec:attack_toy} and case studies in Section~\ref{sec:simulation} using real-world load data reveal the potential vulnerabilities brought by these types of load forecasting attacks.

\subsubsection{Load Maximization}
Under this strategy, the attacker increases the load forecasts as much as possible. Then with an overestimation of the system loads at each time step in the day-ahead stage, the operator tends to turn on more than necessary generation units, which will increases the system operation costs. 

\subsubsection{Load Minimization}
Under this strategy, the attacker decreases the load forecasts as much as possible. Then in day-ahead planning stage, the system operator underestimates the future load, and fewer generators are scheduled than needed. If the real load is not too much higher than the adversarial load, the system can still use spinning reserve to satisfy the underestimated loads, but could cause expensive dispatch. If the real load exceeds the available capacity, load shedding could take place.

\subsection{Toy Example}
\label{sec:attack_toy}
\begin{figure}[h]
	\centering
	\includegraphics[width=1.0 \columnwidth]{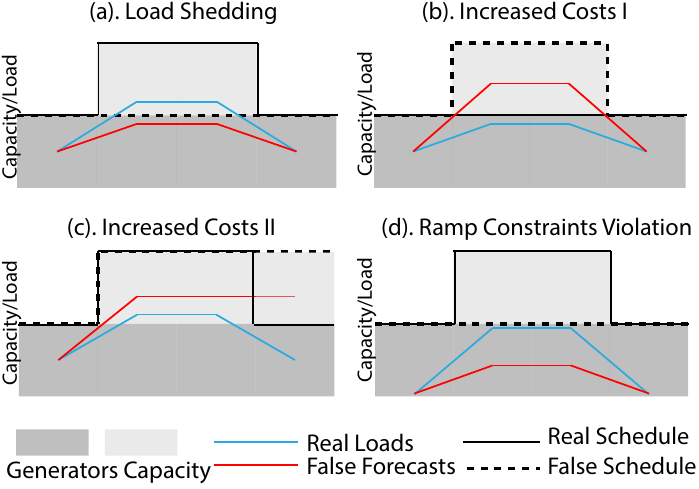}
	\caption{Four cases illustrating consequences by using falsified load forecasts data in a toy example of power system operations. Two generators with same capacity are scheduled for 4 time steps' operations. }
	\label{fig:toy_example}
\end{figure}

To illustrate why such simple attack strategies would cause an increase of system costs and occurrences of operations anomalies, we show an toy example here with $2$ generators of same capacity serving an aggregate load. We demonstrate four possible unexpected cases in Figure \ref{fig:toy_example}. In the simplifying case, we consider a $4$-step forecast and unit commitment. For ease of illustration, we are still assuming there exist ramp constraints and capacity constraints in the toy example, but no minimum up and down time constraints. In Figure \ref{fig:toy_example}(a) and \ref{fig:toy_example}(d), the attacker drives the forecasts lower than the real loads, and we observe either the actual load exceeds the scheduled generator's generation capacity, or actual ramp exceeds the scheduled generator's ramping capacity. In Figure \ref{fig:toy_example}(b) and \ref{fig:toy_example}(c), the attacker either increases the peak load forecasts, or keeps the forecast larger than actual load. Both cases cost the system to keep one more generator online for some time, and dispatch the load in an uneconomical way.

There are other possible attack strategies, such as changing forecasts to random directions, shifting the peak load, cutting the forecast peak load or decreasing the forecasted ramp magnitudes. All these attacks could bring economic and operational damages to the system, but should need more specific design based on the specific load profile, temporal patterns and may require more knowledge of the system.

\subsection{Key Insights}
For the general power system operations including a planning stage~(unit commitment) and a real-time operational stage~(economic dispatch), we observe the following characteristics of impacts by adversarial load forecasts:
\begin{itemize}
	\item Increasing the load maliciously will normally incur extra system costs, such as starting to operate redundant generators, using more expensive generation combinations and etc;
	\item By decreasing the peak load maliciously, system operators would ignore the real peaks of future loads, and schedule fewer generators. This would potentially cause load shedding or failing to follow the severe ramps in the actual load patterns;
	\item We assume an attacker with constrained capability on modifying the input features for load forecast models, and with no knowledge about the system parameters such as generator schedule and load forecasting model parameters. The proposed attack could be even more detrimental if the attacker possesses extra knowledge of the system and implement targeted attack during certain time periods.
\end{itemize}

%% file: simulation.tex
\begin{figure*}[h]
	\centering
	\includegraphics[width=2.1 \columnwidth]{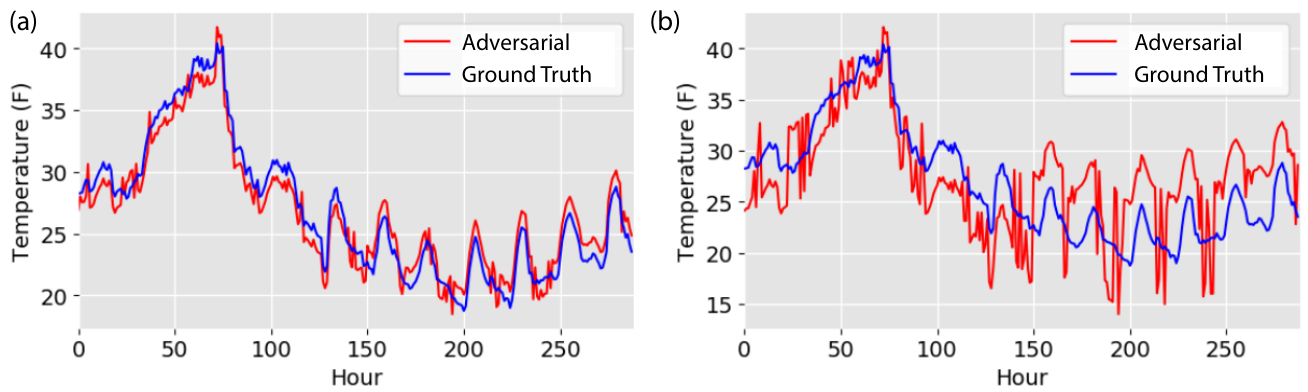}
	\caption{We show $300$ hours of original temperature forecasts and adversarially perturbed temperature using same data as shown in Figure \ref{fig:forecast_example}. (a)Temperature with maximum attack constraint of $1F$; (b). temperature with maximum attack constraint of $5F$.}
	\label{fig:temp}
\end{figure*}

In this section, we show a detailed simulation on real-world Swiss load data, and show the threats posed  by our data injection attacks in several ways. In particular, we first illustrate the proposed attacks could degrade a set of accurate load forecasting algorithms dramatically; we then quantitatively evaluate the damages brought to the system operations, and compare the results with the case using clean data for load forecasting. We demonstrate that attackers with little efforts and knowledge are able to cause load shedding or infeasible dispatch. 

\subsection{Experimental Setup Description}
\textbf{Dataset Description}: We collected and queried hourly actual load data from European Network of Transmission System Operators for Electricity(ENTSO-E)'s API\footnote{https://transparency.entsoe.eu/} ranging from Jan 1st, 2015 to May 16th, 2017, and we followed~\cite{marino2016building} to collect day-ahead historical weather forecasts coming from major cities in Switzerland such as Zurich, Basel, Lucerne and etc. All the weather data were queried from Dark Sky API~\footnote{https://darksky.net/forecast/47.3769,8.5414/us12/en}. We also collect other indicator features $\textbf{X}^{index}$, such as one-hot vectors of hour of day, day of week (weekend or weekday), and season of year. We split $80\%$ of data as our training sets, and use the remaining $20\%$ of data on validating and evaluating the load forecasting prediction accuracy, attack performance and case studies on market operations. Note that even though we collected offline data to train and validate both of our load forecasting and attack models, these data collection procedures could be applied in an online fashion so that attacker could inject real-time adversarial attacks into certain load forecasting models. 

\textbf{Power Systems Setup}: The system has 1 aggregated load for Switzerland based on the ENTSO-E data. The nominal load values are in the range of $[6,500MW,\; 9,500MW]$. We take a simplified power system model of using $7$ generators with total capacity of $11,900MW$, and omit the network constraints. We adopt the generator parameter settings of ramp capacity, generation costs and minimum on/off time based on~\cite{kirschen2018fundamentals}. We set the spinning reserve requirement as 3\% of the total forecasted demand based from~\cite{rebours2005spinning}. During the run of day-ahead unit commitment, either normal day-ahead forecasts or adversarial forecasts are used for generation scheduling; during the run of economic dispatch, the real loads are used for generation dispatch. The models of UC and ED are implemented in Python using PyPSA~\cite{PyPSA}, and these two modules are directly interfaced with the load forecastling and attack algorithms. Note that even in our simiulated system, it ignores the line constraints, while the attacker does not know any information about the system operation, we already observe a set of damages posed by load forecasting attack. We expect more severe effects of attacks with either more generation constraints or less attack constraints.

\textbf{Model Training and Attack Implementation}: We set up three load forecasting models, NN, RNN and LSTM respectively, and use standard stochastic gradient descent methods for model training~\cite{bottou2010large}. For detailed model setup and training, we refer to Appendix.\ref{sec:app}. All three forecast methods could get similar converged validation errors, and as shown in the first column of Table \ref{table:forecast_error}, the errors in mean absolute percentage error~(MAPE) are comparable to the errors reported in several recent studies on load forecasting~\cite{chen2018short, kong2017short}. 
We save the model parameters and keep them away from black-box attackers. For the substitute model training of \texttt{learn and attack} method, we keep the training set $\tilde{\mathcal{D}}$ same as the load forecasting model training set $\mathcal{D}$. Decreasing the size of $\tilde{\mathcal{D}}$ or using different substitute dataset could decrease the performance of \texttt{learn and attack}. We use $L_\infty$ constraints on the attacker's capability \eqref{equ:constraint}, such that the attacker is constrained by the maximum deviation of perturbed temperature values. We validate trained model's performance under attacks with varying constrained values. For details of training techniques, training accuracies, training and attack implementation time, we refer to Appendix \ref{sec:app} and Appendix \ref{sec:app2}.

\begin{table}
	\centering
	\begin{tabular}
		{P{1.8cm}| P{1.8cm}P{1.8cm}P{1.8cm}}
		Forecasts Error (MAPE)&Clean Data&Learn and Attack&Gradient Estimation\\
		\toprule[0.4mm]
		NN & $1.68\%$ & $12.72\%$ & $13.09\%$\\
		RNN & $1.58\%$ &$9.82\%$ &$11.68\%$ \\
		LSTM & $1.51\%$& $9.04\%$&$11.87\%$\\
		\bottomrule[0.4mm]
	\end{tabular}
	\caption{Forecasts errors evaluated on clean test data and adversarial data for 3 different forecast models. Allowed maximum perturbations are $4F$.}
	\label{table:forecast_error}
\end{table}

\begin{figure}[h]
	\centering
	\includegraphics[width=1.0 \columnwidth]{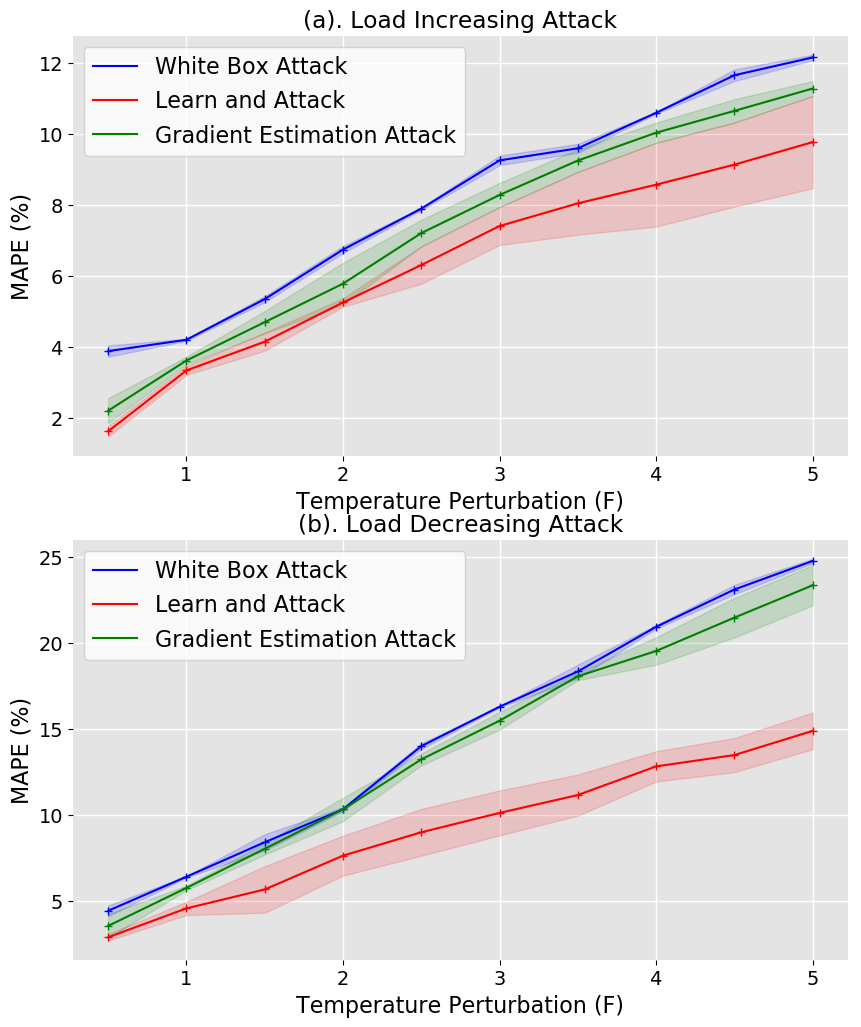}
	\caption{The forecast MAPE under (a). attacks to increase the load; and (b). attacks to decrease the load. Simulations are run for three times with different random seeds, and shaded area denotes the variance.}
	\label{fig:forecast_error}
\end{figure}

\subsection{Load Forecasting Performance}
We calibrate and compare the load forecasting model performance with and without adversarial attacks on test datasets. Though all three models exhibit good performances on clean test data, we inject different level of perturbations generated by \texttt{learn and attack} and \texttt{gradient estimation} methods respectively, and found the forecasting performance decrease drastically as the adversarial perturbations become larger (Table \ref{table:forecast_error}). In Figure ~\ref{fig:forecast_example} we show the RNN's load forecasting results for $300$ hours using \texttt{learn and attack} algorithm with maximum perturbation on temperature of $1F$ and $5F$. The attacker tries to increase the load in the first $150$ hours, and decrease the load in the latter hours. We observe that the algorithm finds the correct attack direction to either increase or decrease the load. What's more, with only $1F$ deviation on temperatures, the load forecasts changes over $500$MW at some time steps. When the attacker increases the perturbation to $5F$, larger forecasts error over $1,200$MW are observed. The temperature profile before and after attack still looks similar, which could avoid system operators' security inspection~(Figure \ref{fig:temp}). Table \ref{table:forecast_error} compares all three load forecasting models' performance using clean and adversarial data. For both \texttt{learn and attack} and \texttt{gradient estimation} algorithms, they distort all three load forecasting models' output and increase model's forecast error. \texttt{Gradient estimation} attack works generally better for all three models, and this is due to estimating the gradients via querying $f_\theta$ directly is more accurate than calculating it from the substitute model and transferring to $f_\theta$.

In Figure~\ref{fig:forecast_error}, we evaluate RNN's load forecasting performance under two attack strategies: load maximization or load minimization. We observe \texttt{gradient estimation} attack causes similar MAPE compared to white box attack.  The load decreasing attack is normally more successful than load increasing attack in terms of MAPE. Load minimization attack is more harmful results than load increasing ones, since increased forecasts only let system operators start up more generations, while adversarially decreasing the forecasted load leads to wrong generation decisions that fails to meet the larger real load.

\subsection{Impact of Attacks on Operation Costs}
\begin{figure}[h]
	\centering
	\includegraphics[width=1.0 \columnwidth]{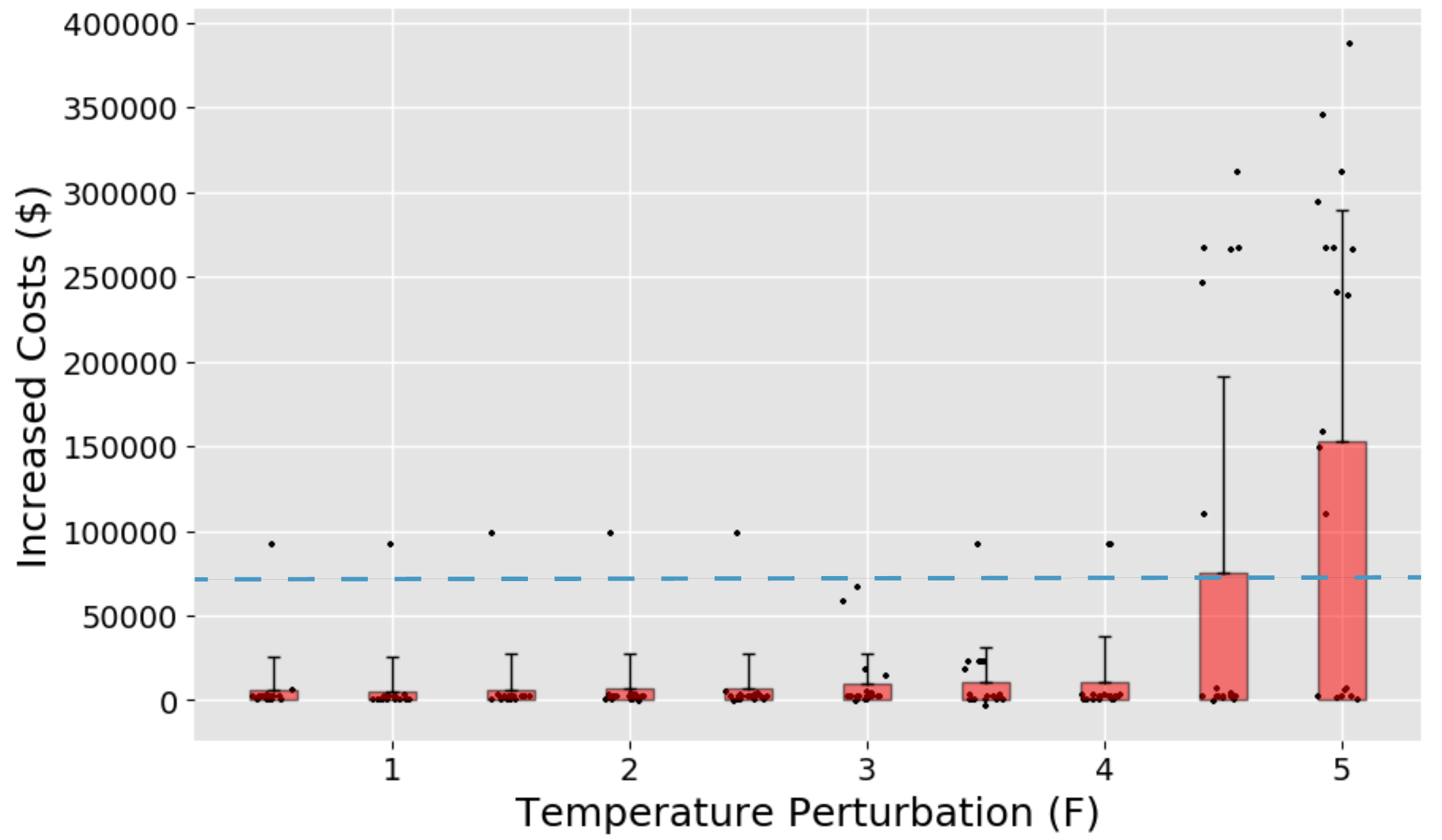}
	\caption{With different level of malicious perturbations injected into temperature features, wrong forecasts increase system operation costs. The nomial hourly operating costs is around \$72,000 (blue dashed line).}
	\label{fig:costs}
\end{figure}

As mentioned earlier, we are interested in the possible consequences caused by wrong forecasts. We first analyze the increased costs caused by adversarial  forecasts. We implement \texttt{learn and attack} algorithm on 3 weeks' random selected test data to increase the forecasted load at each time step. Under such circumstances, the system operator sets day-ahead generator schedule based on adversarial loads larger than actual loads. In Figure~\ref{fig:costs} we show the bar plot of increased costs versus varying perturbation on temperature forecasts. When the temperature perturbations are small, increased costs are limited, and such increments are mostly due to extra start-up costs. When perturbation becomes larger, system operators sometimes derive totally different unit commiment schedule to accomodate higher loads and larger ramps, so in some days we observe larger increase in system costs, of which values are $4-5$ times of nomial hourly operating costs.

\subsection{Impacts of Attacks on Feasibility}
\begin{figure}[h]
	\centering
	\includegraphics[width=1.0 \columnwidth]{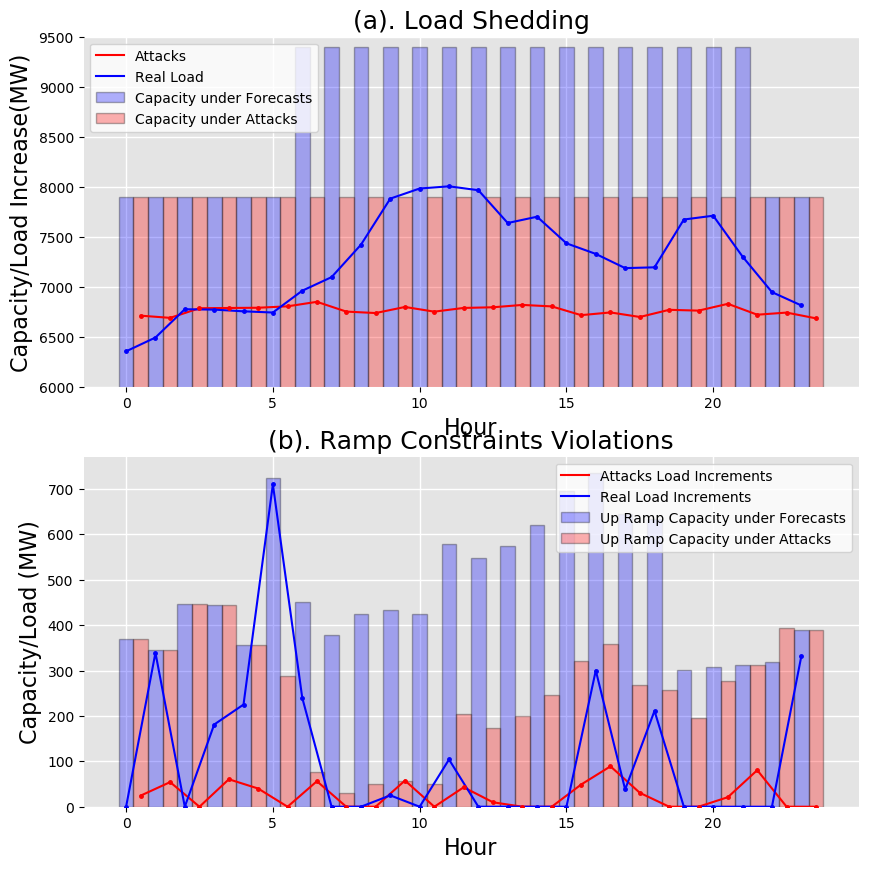}
	\caption{(a). An example showing that forecasts under attack would cause load shedding when real loads exceed total generation capacity; bars indicate generators' total capacity. (b). An example showing that forecasts under attack would cause violation on ramp constraints during economic dispatch; bars indicate generators' available total up-ramp capabilities. Maximum allowed perturbations are $4F$.}
	\label{fig:UC}
\end{figure}

In addition to increasing system costs, adversarial attacks on load forecasting could even lead to worse situations. We illustrate a load minimization strategy that leads to infeasible solutions~(e.g., load shedding, ramp constraint violations) to the economic dispatch problem. We implement both \texttt{learn and attack} and \texttt{gradient estimation} algorithms with maximum perturbation of $4F$, and test the results on $10$ weeks' load data. In Table \ref{table:anomaly}, we note the occurrence frequencies of both load shedding and ramp constraints violation. Since $4F$ change in temperature forecasts can lead to over $1,000$MW decreasing on load forecasts, system operators tend to keep fewer generators on. This leads to many days' generation capacity fall short of the load, and the scheduled generators can not fulfill the large ramps in real load profiles. In Figure \ref{fig:UC} we show two examples on this two kind of failures respectively. In Figure~\ref{fig:UC}(a), during peak hours, the adversarial load forecasts let the system operator schedule one $1,500MW$ generator off compared to the case of correct forecasts. Even taking the spinning reverse during the day-ahead unit commitment, the actual load at the mid of the day exceeds the adversarial load by over $1,000MW$ and the total load exceeds the generator capacity. In Figure~\ref{fig:UC}(b), the actual loads are increasing rapidly at hour 5 and 6, yet the adversarials load profile flattens such ramps, and cause the scheduled generators incapable of meeting the large ramp. We expect more frequent violations of ramp constraints if the attacker specifically design attack strategies based on the load patterns.

\begin{table}
	\centering
	\begin{tabular}
		{P{3.8cm}| P{1.8cm}P{1.8cm}}
		Occurrences  
		\quad \quad \quad \quad \quad (Number of Days) &Learn and Attack&Gradient Estimation\\
		\toprule[0.4mm]
		Load Shedding & $27.14\%$ & $31.43\%$  \\
		Ramp Constraints Violation& $90.0\%$ &$85.71\%$  \\
		\bottomrule[0.4mm]
	\end{tabular}
	\caption{The occurrence frequencies of load shedding and ramp constratins violation in 10 weeks' system operation test. Both attackers are allowed to inject $4F$ perturbations on temperature features.}
	\label{table:anomaly}
\end{table}

%% file: conclusion.tex
In this paper, we studied the potential vulnerabilities generally existing in many load forecasting algorithms. Such vulnerabilities have been overlooked by the development of many forecasting techniques. We design two attack algorithms which do not require much knowledge about the forecast algorithms, but lead to large increase in forecast errors with adversarial data injections in load forecasting input features. The proposed attack adversarilly manipulate the load forecasting values either to  increase or decrease, and thus provide system operators wrong information on future demands. Experiments on real-world load datasets demonstrate such threats over power system operations. Such threats model along with damage analysis indicate that there need more security evaluations in the design and implementation of load forecasting algorithms. In order to mitigate the damages brought by such false data injection attacks, countermeasures in building robust load forecasting algorithms are strongly recommended, which may include anomaly detection techniques considering input data distribution as well as other robust statistics.

%% file: appendix.tex
\section {Details on Load Forecasting Algorithms}
\label{sec:app}
We set up all load forecasting models using Tensorflow~\cite{abadi2016tensorflow} package in Python. Standard model architectures such as Dropout layers and nonlinear activation functions (e.g., ReLU or Sigmoid functions) are adopted in the deep learning models~\cite{nair2010rectified}. Since all three networks are set up to solve the load forecasting regression problem, we set the first layer having most neurons, and decrease the number of units in subsequent layers.

\begin{table}[h]
	\centering
\begin{tabular}{>{\centering}m{3.8cm} >{\centering}m{0.8cm} >{\centering}m{0.8cm} >{\centering\arraybackslash}m{0.8cm}}
		Forecasts Models&NN&RNN&LSTM\\
		\toprule[0.4mm]
		Number of Layers & 4 & 3 & 3\\
		Training Epochs & 20 &30 &30 \\
		Hidden Units in First Layer& 512& 64&64\\
		\bottomrule[0.4mm]
	\end{tabular}
	\caption{Model architectures and training configurations for load forecasting algorithms used in the simulations.}
	\label{table:nn}
\end{table}

\begin{figure}[h]
	\centering
	\includegraphics[width=1.0 \columnwidth]{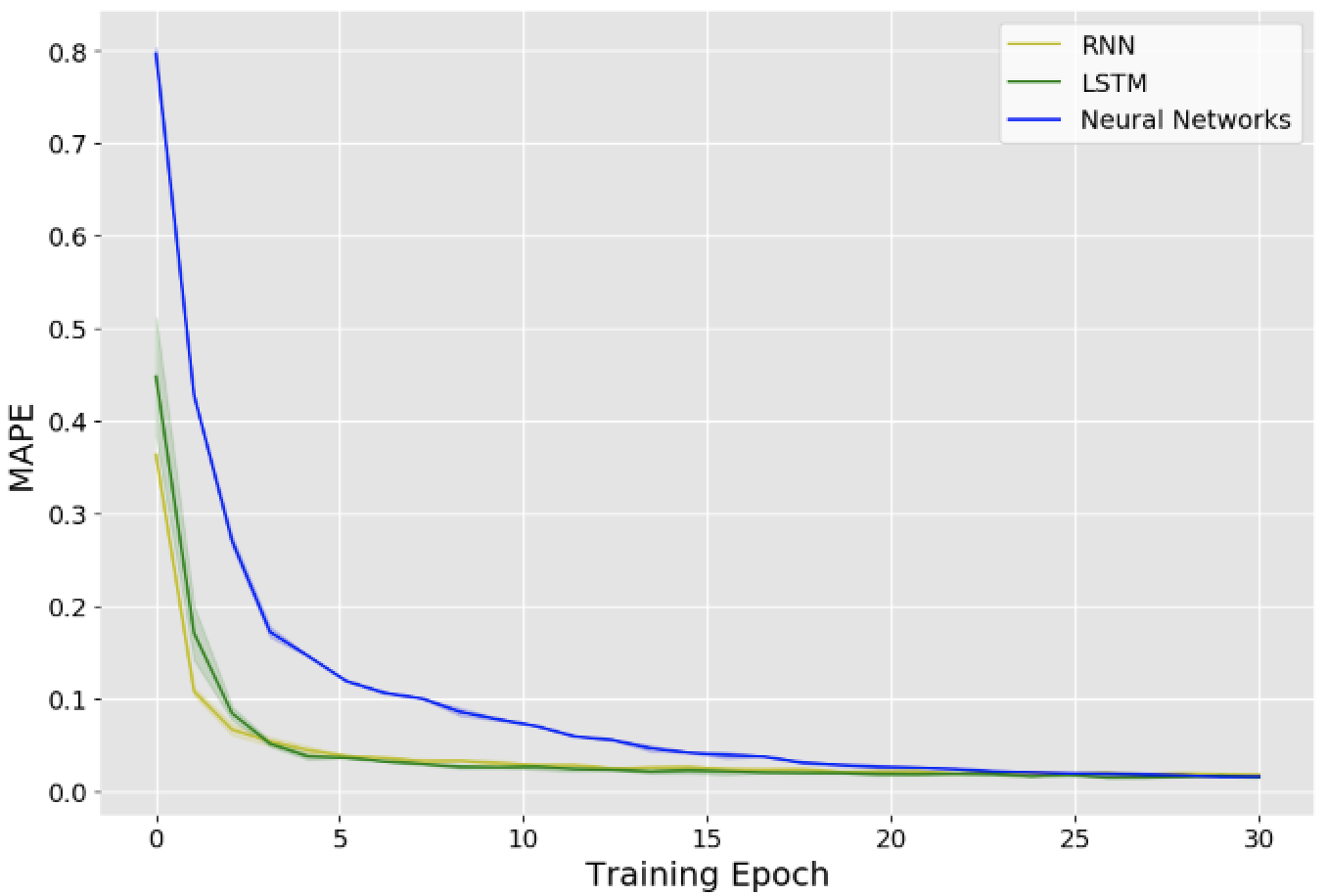}
	\caption{All three forecasting models, show convergence of forecast error on validation data as  training evolves. Shaded areas show the variance of MAPE.}
	\label{fig:forecast_trainingr}
\end{figure}

As shown in Figure \ref{fig:forecast_trainingr}, all three model's loss are converged during training, and we use the trained model in the subsequent planning and operation problem as well as the testbed for attack algorithms. Plots are showing the mean and variance during 3 runs with different random seeds.

\section{Computation Time}
\label{sec:app2}
We recorded the computation time for neural network training and the implementation time for two proposed attack algorithms. All time are recorded on a laptop with Intel 2.3GHz Core i5-8259U 4 Cores CPU and 8 GB RAM. The training time for NN, RNN and LSTM are calculated for 20, 30 and 30 epochs respectively. The implementation time for the attacks are averaged over all test instances. We could observe that \texttt{learn and attack} approach takes longer time than  \texttt{gradient estimation} due to the longer time taken to calculate gradient signs over the whole neural networks; and as LSTM includes more complicated model architectures, it takes longer time to find the adversarial instance. Yet compared to the long model training time and application scenarios of day-ahead forecasts, the attacker is still efficient enough to find the adversarial perturbations.

\begin{table}[h]
	\centering
	\begin{tabular}{>{\centering}m{4.8cm} >{\centering}m{0.8cm} >{\centering}m{0.8cm} >{\centering\arraybackslash}m{0.8cm}}
		Forecasts Models&NN&RNN&LSTM\\
		\toprule[0.4mm]
		Training Time & 12.988 & 47.998 & 143.830\\
		\texttt{Learn and Attack} & 0.133 &0.157 &0.579 \\
		\texttt{Gradient Estimation Attack} & 0.082 & 0.119 &0.253\\
		\bottomrule[0.4mm]
	\end{tabular}
	\caption{Computation time (in seconds) for load forecasting model training and implementation time for attacks.}
	\label{table:time}
\end{table}